\documentclass[aps, prd, preprint, superscriptaddress, nofootinbib, amsfonts, amssymb, amsmath, longbibliography, showkeys]{revtex4-1}
\usepackage{graphicx, bm, color, wrapfig}
\usepackage[colorlinks, linkcolor=blue, anchorcolor=blue, citecolor=blue, urlcolor=black, breaklinks=true]{hyperref}
\allowdisplaybreaks

\begin{document}

\title{Globally Stable Dark Energy in \texorpdfstring{$F(R)$}{F(R)} Gravity}

\author{Hua Chen}
\email{huachen@mails.ccnu.edu.cn}
\affiliation{Institute of Astrophysics, Central China Normal University, Wuhan 430079, China}

\begin{abstract}
    $F(R)$ models for dark energy generally exhibit a weak curvature singularity, which can be cured by adding an $R^2$ term.
    This correction allows for a unified description of primordial and late-time accelerated expansions.
    However, most existing models struggle to achieve this, as they become unstable over certain negative ranges of the Ricci scalar,
    where either the first or second derivative of $F(R)$ turns negative.
    These instabilities may disrupt the post-inflationary evolution when the Ricci scalar oscillates about the vacuum state after the $R^2$ inflation.
    In this paper, we introduce a model-building process to guarantee global stability, i.e., the first and second derivatives are positive for all real Ricci scalars.
    By extending the idea from Appleby and Battye, we demonstrate that viable models can be constructed by imposing a positive, bounded first derivative of $F(R)$ with a sigmoid shape.
    Building upon this framework, we first reformulate and generalize the original Appleby-Battye model.
    Then, we propose a new dark energy model that successfully explains the acceleration of cosmic expansion and passes local gravity tests.
\end{abstract}


\maketitle

\section{Introduction}

Since the discovery of the accelerated expansion of the Universe about a quarter-century ago \cite{SupernovaSearchTeam:1998fmf, SupernovaCosmologyProject:1998vns}, numerous attempts have been made to explain the underlying repulsive energy source, referred to as dark energy \cite{Li:2012dt, Yoo:2012ug, Mortonson:2013zfa, Klimchitskaya:2024dvk}.
In addition to the static cosmological constant \cite{Planck:2018vyg, Bull:2015stt, Perivolaropoulos:2021jda}, dark energy could be attributed to a dynamical scalar field emerging, for instance, from $F(R)$ gravity \cite{Sotiriou:2008rp, DeFelice:2010aj, Nojiri:2017ncd}.
By replacing the Ricci scalar $R$ with a general function $F(R)$, $F(R)$ gravity introduces an additional scalar degree of freedom, dubbed scalaron.

Capozziello was the first to explore cosmic acceleration within the framework of $F(R)$ gravity by incorporating quintessence,
called curvature quintessence \cite{Capozziello:2002rd}.
Subsequently, Carroll et al. proposed the first well-studied model \cite{Carroll:2003wy}, while Nojiri et al. proposed the first potential unified model to describe both primordial and late-time acceleration within a single $F(R)$ Lagrangian \cite{Nojiri:2003ft}.
Such a light scalar field would mediate a long-range fifth force, which is tightly constrained by solar-system tests through the Parameterized Post-Newtonian (PPN) parameter.
Fortunately, nonlinear effects can screen the propagation of the fifth force via the chameleon mechanism \cite{Khoury:2003aq, Khoury:2003rn, Brax:2004qh}.
After a few years of exploration, several viable models have been successfully developed \cite{Hu:2007nk, Starobinsky:2007hu, Appleby:2007vb, Tsujikawa:2007xu, Cognola:2007zu}.

However, it was soon recognized that these models typically lead to a weak curvature singularity, occurring in the past \cite{Starobinsky:2007hu, Tsujikawa:2007xu, Frolov:2008uf, Appleby:2008tv} (for discussions on finite-time singularities, see \cite{Bamba:2008ut, Abdalla:2004sw, Nojiri:2008fk, Capozziello:2009hc}).
Consequently, the scalaron mass exceeds the Planck mass even during the matter-dominated era, signaling the breakdown of $F(R)$ gravity as an effective field theory.
It has been found that adding an $R^2$ term can naturally remove the singularity, and the resulting $R^2$-corrected dark energy may provide a unified description of primordial and late-time acceleration of the cosmic expansion \cite{Kobayashi:2008wc, Appleby:2009uf, Elizalde:2010ts, Lee:2012dk}.
For general considerations or related work, see Refs.~\cite{Nojiri:2006gh, Nojiri:2007as, Nojiri:2007cq, Nojiri:2010wj, Odintsov:2019evb, Katsuragawa:2017wge, Chen:2019kcu, Chen:2022zkc}.

In such a unified scenario, the standard sequence of inflation, reheating, radiation-dominated, matter-dominated, and dark energy-dominated epochs must be preserved.
In the original $R^2$ inflationary model, post-inflationary reheating begins when the linear $R$ term dominates over the nonlinear $R^2$ term and oscillates around the vacuum state $R=0$.
Adhering to this reheating mechanism suggests that not all viable dark energy models can fulfill these criteria, as several become unstable when $R<0$.
The only viable model, to the best of my knowledge, is the Appleby-Battye model \cite{Appleby:2009uf}, which, unlike others, was specifically designed with the condition $F_R>0$ for all ranges of the Ricci scalar \cite{Appleby:2007vb}.
For recent cosmological constraints, see \cite{Ribeiro:2023yhh}.

In this study, we will generalize the idea from the Appleby-Battye model and demonstrate that the favored form of $F_R(R)$ is a bounded function with a sigmoid shape, with its midpoint (or center of symmetry) determined by the critical curvature.
In addition, the $\Lambda$CDM model (for positive $R$) can be regarded as an extreme case of these sigmoid functions, i.e., the Heaviside delta function of $R$.

The Appleby-Battye model contains exponential terms, and as we will show, the first derivative is the logistic function.
For comparison, we will also introduce a new model whose first derivative approximates a power-law behavior in the high-curvature limit.
Consequently, our model resembles the well-studied Hu-Sawicki model in the high-curvature limit but achieves global stability.
We demonstrate that this model can account for the current cosmic acceleration by explicitly solving the Friedmann equation.
Furthermore, the effective Equation of State (EoS) parameter for the dark energy component exhibits phantom crossing behavior, a common feature of $F(R)$ dark energy models \cite{Amendola:2007nt, Bamba:2010iy, Bamba:2010zxj}.
Subsequently, we constrain the model parameters by confronting our model with local gravity tests.
Finally, we discuss the unified description of dark energy and inflation within the framework of globally stable models.

Throughout this paper, we adopt natural units with $c=\hbar=1$ and focus exclusively on the metric formalism, where the metric signature is $(-,+,+,+)$.
The prime ``$\prime$" and subscript ``$R$" denote derivatives with respect to the function's argument and Ricci scalar, respectively.

\section{Basis of \texorpdfstring{$F(R)$}{F(R)} gravity}

The action of $F(R)$ gravity is given by
\begin{equation}
    \label{eq:JF_action}
    S=\frac{1}{2\kappa^{2}}\int\,d^{4}x\sqrt{-g}F(R)+S_{\mathrm{Matter}}\,,
\end{equation}
where $\kappa^{2}\equiv8\pi G\equiv1/M_{\mathrm{Pl}}^{2}$ with $G$ being the bare gravitational constant and $M_{\mathrm{Pl}}\simeq 2.44\times10^{18}\,\mathrm{GeV}$ being the reduced Planck mass,
and $S_{\mathrm{Matter}}$ is the matter action describing all non-gravitational parts.
For convenience, we define $F(R)\equiv R+f(R)$, such that the modification to the general relativity is encoded in $f(R)$.

The variation with respect to the metric $g^{\mu\nu}$ gives the modified field equation,
\begin{equation}
    \label{eq:JF_field_eq}
    g_{\mu\nu}\Box F_{R}-\nabla_{\mu}\nabla_{\nu}F_{R}+R_{\mu\nu}F_{R}-\frac{1}{2}g_{\mu\nu}F=\kappa^{2}T_{\mu\nu}\,.
\end{equation}
where $F_{R}\equiv\partial F/\partial R$.
The trace of the field equation \eqref{eq:JF_field_eq} reads
\begin{equation}
    \label{eq:trace_eq}
    3\square f_{R}+Rf_{R}-2f-R=\kappa^{2}T_{\mu}^{\mu}\,,
\end{equation}
indicating a propagating scalar degree of freedom, $f_{R}$, dubbed scalaron.
The dynamics of the scalaron is subject to an effective potential $U_{\mathrm{eff}}$:
\begin{equation}
    \square f_{R} =\frac{\partial U_{\mathrm{eff}}}{\partial f_{R}}=\frac{1}{3}\left(2f-Rf_{R}+R+\kappa^{2}T_{\mu}^{\mu}\right)\,.
\end{equation}
The mass squared of the scalaron at the extremum $\frac{\partial U_{\mathrm{eff}}}{\partial f_{R}}=0$ is defined by
\begin{equation}
    m_{f_{R}}^{2}=\frac{\partial^{2}U_{\mathrm{eff}}}{\partial f_{R}^{2}}=\frac{1}{3f_{RR}}\left(1+f_{R}-Rf_{RR}\right)\,.
\end{equation}

The vacuum solutions can be determined from
\begin{equation}
    \label{eq:vacuum_equation}
    2f - Rf_{R} + R=0\,.
\end{equation}
Notably, we have $R=4\Lambda$ in the $\Lambda$CDM model.
If the solution satisfies
\begin{equation}
    \label{eq:stable_de_Sitter}
    \frac{1+f_{R}}{f_{RR}}>R\,,
\end{equation}
then the de Sitter point is future-stable with the mass squared being positive.

We can rewrite the action in the form of scalar-tensor theories by introducing an auxiliary field $\chi$:
\begin{equation}
    S=\frac{1}{2\kappa^{2}}\int d^{4}x\,\sqrt{-g}\left[F(\chi)+F_{,\chi}(\chi)\left(R-\chi\right)\right]+S_{\mathrm{Matter}}\,.
\end{equation}
The variation of the action with respect to $\chi$ yields
\begin{equation}
    F_{,\chi\chi}(\chi)(R-\chi)=0\,.
\end{equation}
If $F_{,\chi\chi}$ is not identically vanishing, then $\chi=R$.
Therefore, we can define
\begin{align}
    \psi    & \equiv F_{,\chi}(\chi)\,,                                                        \\
    U(\psi) & \equiv\frac{\chi F_{,\chi}(\chi)-F(\chi)}{2\kappa^{2}}\,,\label{eq:jf_potential}
\end{align}
such that
\begin{equation}
    \label{eq:Brans-Dicke_form}
    S=\int d^{4}x\,\sqrt{-g}\left(\frac{1}{2\kappa^{2}}\psi R-U(\psi)\right)+S_{\mathrm{Matter}}\,.
\end{equation}

As the field equation \eqref{eq:trace_eq} is a fourth-order differential equation of $g_{\mu\nu}$,
an intuitive way is to transform the action into the Einstein frame through the Weyl transformation:
\begin{equation}
    \tilde{g}_{\mu\nu}=F_{R}(R)g_{\mu\nu}\equiv e^{-2\beta\kappa\varphi}g_{\mu\nu}\,,
\end{equation}
where $\beta$ is the coupling constant, and $\varphi$ is a new scalar field defined in the Einstein frame:
\begin{equation}
    \label{eq:ef_field}
    \varphi\equiv-\frac{1}{2\beta}M_{\mathrm{Pl}}\ln F_{R}\,.
\end{equation}
Choosing $\beta=-\frac{1}{\sqrt{6}}$, the action is rewritten as the Einstein-Hilbert term plus a canonical scalar field as follows:
\begin{equation}
    \label{eq:ef_action}
    \tilde{S}=\int d^{4}x\,\sqrt{-\tilde{g}}\left[\frac{1}{2\kappa^{2}}\tilde{R}-\frac{1}{2}\tilde{g}^{\mu\nu}(\tilde{\partial}_{\mu}\varphi)(\tilde{\partial}_{\nu}\varphi)-V(\varphi)+e^{4\beta\kappa\varphi}\mathcal{L}_{\mathrm{Matter}}\right]\,,
\end{equation}
where $V(\varphi)$ is the potential of the scalaron given by
\begin{equation}
    \label{eq:ef_potential}
    V(\varphi)=\frac{1}{2\kappa^{2}}\frac{RF_{R}(R)-F(R)}{F_{R}^{2}(R)}=\frac{U(\psi)}{F_{R}^{2}(R)}\,.
\end{equation}

In the Einstein frame, the scalaron is non-minimally coupled to the matter fields, and the energy-momentum tensor of the matter fields is
\begin{equation}
    \tilde{T}_{\nu}^{\mu}=\tilde{g}^{\mu\rho}\tilde{T}_{\rho\nu}=e^{4\beta\kappa\varphi}T_{\nu}^{\mu}\,.
\end{equation}
Similarly, the energy density $\tilde{\rho}=e^{4\beta\kappa\varphi}\rho$ and pressure $\tilde{P}=e^{4\beta\kappa\varphi}P$.
Since $w=P/\rho$, the EoS parameter is independent of frame transformation.
On the other hand, the conserved energy density in the Einstein frame is
\begin{equation}
    \label{eq:ef_conserved_density}
    \tilde{\rho}_{*}\equiv e^{-\left(1-3w\right)\beta\kappa\varphi}\tilde{\rho}=e^{3\left(1+w\right)\beta\kappa\varphi}\rho\,,
\end{equation}
rather than $\tilde{\rho}$.
For dust-like matter $w=0$, $\tilde{\rho}_{*}=e^{3\beta\kappa\varphi}\rho$; for radiation, $w=1/3$, $\tilde{\rho}_{*}=e^{4\beta\kappa\varphi}\rho=\tilde{\rho}$.

The equation of motion for the scalaron is
\begin{equation}
    \label{eq:EF_eom}
    \tilde{\square}\varphi=V_{,\varphi}-\beta\kappa e^{4\beta\kappa\varphi}T_{\mu}^{\mu}\,.
\end{equation}
Thus, the scalaron is subject to an effective potential:
\begin{equation}
    \label{eq:effective_potential}
    V_{\mathrm{eff}}(\varphi)=V(\varphi)+e^{\left(1-3w\right)\beta\kappa\varphi}\tilde{\rho}_{*}\,.
\end{equation}
Accordingly, the effective mass squared is defined by
$m_{\varphi}^{2}\equiv V_{\mathrm{eff},\varphi\varphi}(\varphi_{\min})$,
where $\varphi_{\mathrm{min}}$ is the equilibrium value corresponding to the minimum of the effective potential, which is determined by the trace $T_{\mu}^{\mu}$.

\section{Dark energy}

Generally, a cosmologically viable model for dark energy in $F(R)$ gravity should at least satisfy the following viability conditions \cite{Amendola:2007nt, Appleby:2009uf}:
\begin{equation}
    F_{R}(R)>0,\quad F_{RR}(R)>0\,\qquad \text{for}\ R\geq R_{1}\,,
\end{equation}
where $R_{1}$ is the maximal de Sitter solution.
The first condition is imposed to evade anti-gravity and the ghost state of the graviton \cite{Nariai:1973, Gurovich:1979xg, Appleby:2009uf}, and the second condition ensures that the mass squared is positive and avoid the Dolgov-Kawasaki and tachyon instabilities \cite{Dolgov:2003px, Song:2006ej, Amendola:2006kh, Amendola:2006we, Appleby:2009uf}.
Other viable conditions can be inferred from dynamical analysis \cite{Amendola:2007nt}.
To be distinguished from the $\Lambda$CDM model, it is suggested that $F(R)$ gravity does not contain a true cosmological constant, i.e., $F(0)=0$ \cite{Starobinsky:2007hu}.

\begin{figure}
    \centering
    \includegraphics[scale=0.6]{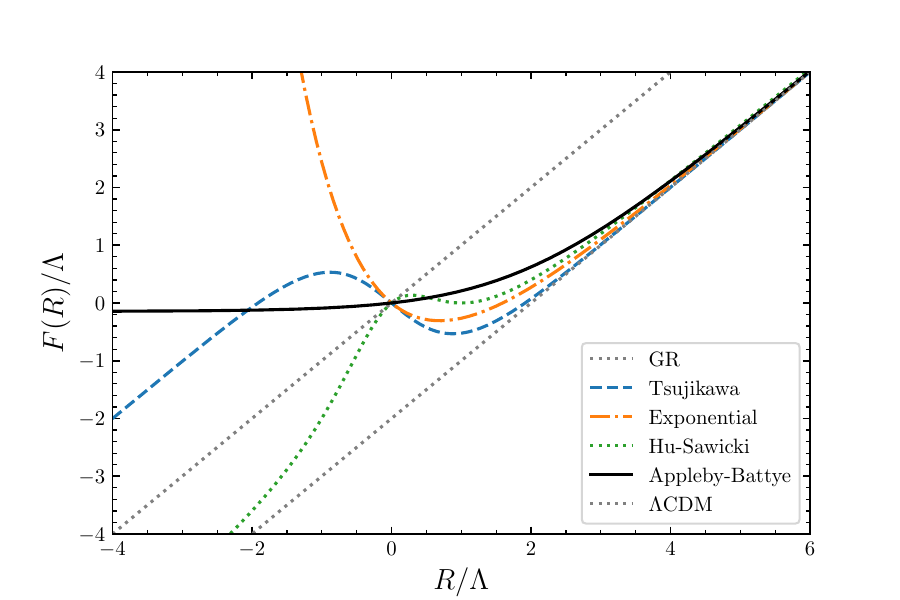}
    \caption{Comparison of $F(R)/\Lambda$ versus $R/\Lambda$ for existing models, with all the parameters set to unity. The Starobinsky model is not shown here, as it coincides with the Hu-Sawicki model for $n=1$.}
    \label{fig:F_cmp}
\end{figure}

The above conditions impose stringent constraints on possible forms of the $F(R)$ function.
Some well-studied viable models satisfying the above requirements are (with different notations to the original papers)
\begin{itemize}
    \item Hu-Sawicki model \cite{Hu:2007nk}
          \begin{equation}
              F(R)=R-R_{\mathrm{c}}\frac{c_{1}\left(R/R_{\mathrm{c}}\right)^{2n}}{c_{2}\left(R/R_{\mathrm{c}}\right)^{2n}+1}\,,
          \end{equation}
          where $c_{1},c_{2}>0$ are free parameters. $R_{\mathrm{c}}\equiv\frac{\kappa^{2}\bar{\rho}_{\mathrm{m}0}}{3}$, where $\bar{\rho}_{\mathrm{m}0}=\bar{\rho}_{\mathrm{m}}(\ln a=0)$ is the current averaged matter density.
    \item Starobinsky model \cite{Starobinsky:2007hu}
          \begin{equation}
              F(R)=R-\lambda R_{\mathrm{c}}\left[1-\left(1+\frac{R^{2}}{R_{\mathrm{c}}^{2}}\right)^{-n}\right]\,,
          \end{equation}
          where $n,\lambda>0$, and $R_\mathrm{c}$ is the order of the observed cosmological  constant $\Lambda\simeq10^{-66}\,\mathrm{eV^{2}}$.
    \item Tsujikawa model \cite{Tsujikawa:2007xu}
          \begin{equation}
              F(R)=R-\lambda R_{\mathrm{c}}\tanh\left(\frac{R}{R_{\mathrm{c}}}\right)\,,
          \end{equation}
          where $\lambda,b>0$.
    \item Exponential model \cite{Cognola:2007zu}
          \begin{equation}
              F(R)=R-\lambda R_{\mathrm{c}}\left(1-e^{-R/R_{\mathrm{c}}}\right)\,,
          \end{equation}
          where $\lambda, b>0$.
\end{itemize}
These models are carefully constructed by requiring the $F(R)$ to satisfy the viability conditions.
Physically, it is required that $|F(R)-R|\ll R$, $F_{R}-R\ll 1$, and $RF_{RR}\ll 1$ for $R\gg R_{0}=R(t_0)$ to preserve the success of general relativity \cite{Appleby:2009uf}.
Practically, as shown in Fig. \ref{fig:F_cmp}, all the aforementioned models approximate to $F(R)\approx R-\lambda R_\mathrm{c}+\cdots$ in the high-curvature limit.
Therefore, $\lambda R_\mathrm{c}$ can be interpreted as $2\Lambda$.
Although it is claimed that there is no true cosmological constant in $F(R)$ gravity if we impose $F(0)=0$,
this effectively introduces a cosmological constant in the Lagrangian.
Mathematically, this is because one has to introduce a critical curvature $R_\mathrm{c}$ to ensure the correct normalized argument,
$R/R_\mathrm{c}$, in the nonlinear function $f(R)$.
This leads to different asymptotic behaviors in various limits, with $F(R\gg R_\mathrm{c})\to R-\lambda R_\mathrm{c}$ in the high-curvature limit.
In this study, we adopt the asymptotic behavior as a constraint for parameterizing models while acknowledging that some alternative models may not satisfy this condition.
However, this behavior is not a necessary condition for our model-building process.

\begin{figure}
    \centering
    \includegraphics[width=\textwidth]{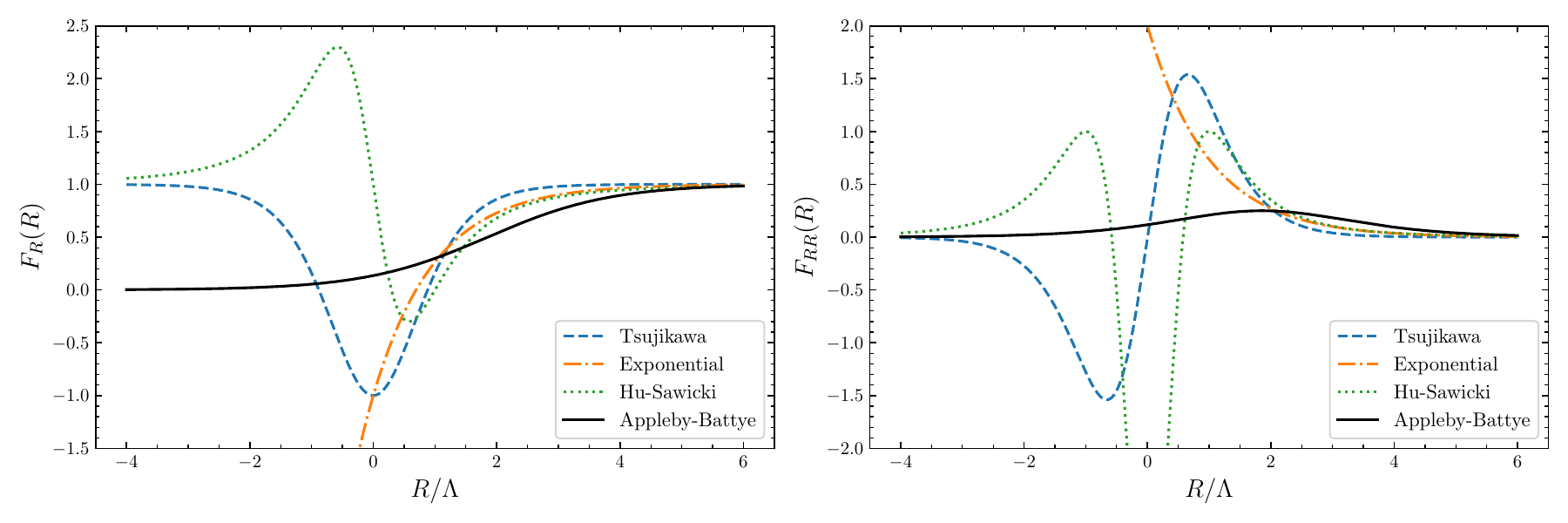}
    \caption{Comparison of $F_{R}(R)$ (left panel) and $\Lambda F_{RR}(R)$ (right panel) versus $R/\Lambda$ for existing models, with all the parameters set to unity.}
    \label{fig:FR_FRR_cmp}
\end{figure}

As shown in Fig. \ref{fig:FR_FRR_cmp}, $F_R$ and $F_{RR}$ become negative when $R \sim 0$ for most models listed above.
Although this does not pose an immediate issue as $R \sim 0$ is generally inaccessible in the current universe, it introduces challenges when attempting to build a unified model by adding an $R^2$ term to the $F(R)$ Lagrangian to explain both primordial and late-time cosmic acceleration.
Specifically, in the original $R^2$ inflation scenario, efficient reheating necessitates an oscillating $R$ around the vacuum state $R = 0$.
Consequently, for model stability, it is essential that $F_{R} > 0$ and $F_{RR} > 0$ hold within the accessible negative $R$ region.
Otherwise, for instance, if $F_{RR} < 0$, the vacuum state becomes unstable, and the scalaron mass diverges when $F_{RR} = 0$ \cite{DeFelice:2010aj}.

In contrast, the Appleby-Battye model (the black solid line in Fig. \ref{fig:FR_FRR_cmp}) has no such problem, as it was constructed by initially imposing $F_R>0$ for all $R$ \cite{Appleby:2007vb}.
The advantage of constructing models starting from $F_{R}(R)$ rather than $F(R)$ is that it is easier to ensure $F_{R}>0$ and $F_{RR}>0$ for all $R$.
However, the disadvantage is that it is not always straightforward to explicitly derive the primitive $F(R)$ function by integrating the given $F_{R}$ function.
$F_{RR}>0$ indicates that $F_{R}$ should be monotonically increasing.
In addition, $F_{R}>0$ and $F(R\gg R_\mathrm{c})\to R-2\Lambda$ require that it be a bounded function, $0<F_{R}<1$ \cite{Appleby:2007vb}.
The first derivative of the Appleby-Battye model \eqref{eq:ab_original} was chosen as  \cite{Appleby:2007vb}
\begin{equation}
    \label{ab_FR}
    F_{R} =\frac{1}{2}\left(1+\tanh(aR-b)\right)\,,
\end{equation}
where $a$ and $b$ are free parameters.
The resulting $F(R)$ model reads~\cite{Appleby:2007vb}
\begin{equation}
    \label{eq:ab_original}
    F(R) = \frac{1}{2}R + \frac{1}{2a}\ln\left[\cosh(aR)-\tanh(b)\sinh(aR)\right]\,.
\end{equation}
The other form often seen in the literature is written as \cite{Appleby:2009uf}
\begin{equation}
    F(R)=\frac{R}{2}+\frac{\epsilon_{\mathrm{AB}}}{2}\ln\left[\frac{\cosh\left(R/\epsilon_{\mathrm{AB}}-b\right)}{\cosh(b)}\right]\,,
\end{equation}
where $\epsilon_{\mathrm{AB}}=R_{\mathrm{vac}}/\left(b+\ln(2\cosh b)\right)$, with $b$ being the model  parameter and $R_{\mathrm{vac}}\sim 4\Lambda$ being the vacuum curvature.

Eq.~\eqref{ab_FR} is not the only choice that satisfies $0<F_R<1$ and $F_{RR}>0$ for all $R$.
To facilitate finding the primitive function for a given $F_{R}(R)$, we suggest that the desired $F_{R}(R)$ function should adopt a sigmoid shape, with its midpoint (or center of symmetry) determined by the critical curvature $R_\mathrm{c}$.
Moreover, the interval over which $F_{R}(R)$ transitions from $F_{R}(R)=0$ to $F_{R}(R)=1$ should be narrow, enabling $F_{R}(R)$ to approach the $\Lambda$CDM limit rapidly; here, (the positive part of) the $\Lambda$CDM is regarded as an extreme case of the sigmoid function, i.e., the Heaviside delta function.
Then, the $F(R)$ model can be derived by integrating the given $F_{R}$ function and fixing the integration constant through the condition $F(0)=0$.
We shall first reformulate the original Appleby-Battye model \eqref{eq:ab_original} to show the exponential suppressed term explicitly.
Then, we propose a new model whose first derivative exhibits a power-law behavior in the high-curvature limit.

\subsection{Appleby-Battye model revisited}

We note that the function in Eq. \eqref{ab_FR} is the well-known logistic function:
\begin{equation}
    \label{eq:logistic_function}
    S(x)=\frac{L}{1+e^{-k(x-x_{0})}}\,,
\end{equation}
where $L$ is the maximum, $k$ controls the growth rate or the steepness of the curve, and $x_{0}$ is the midpoint.
Analogously, the suggested form of the $F_{R}$ function is
\begin{equation}
    \label{eq:logistic_fr}
    F_{R}(R)=\frac{1}{1+be^{-R/R_{\mathrm{c}}}}\,.
\end{equation}
One notes that the midpoint is $\left(R_{\mathrm{c}}\ln b,\frac{1}{2}\right)$, where the parameter $b$ controls the shift of the midpoint, and the curvature scale $R_{\mathrm{c}}$ controls the steepness of the curve or the clockwise rotation with respect to the midpoint.
Specifically, for larger values of $R_{\mathrm{c}}$, the curve becomes shallower; conversely, for smaller values of $R_{\mathrm{c}}$, the curve becomes steeper, and the function approaches the $\Lambda$CDM limit with $F_{R}=1$.

Integrating $F_R(R)$ with respect to $R$ yields the primitive function,
\begin{equation}
    F(R) =R_{\mathrm{c}}\ln\left(e^{R/R_{\mathrm{c}}}+b\right)+C\,,
\end{equation}
where $C$ is an integration constant.
It can be fixed by imposing $F(0)=0$, such that
\begin{equation}
    C=-R_{\mathrm{c}}\ln\left(1+b\right)\,.
\end{equation}
Then, the full $F(R)$ function reads
\begin{equation}
    F(R) =R-R_{\mathrm{c}}\ln\left(1+b\right)+R_{\mathrm{c}}\ln\left(1+be^{-R/R_{\mathrm{c}}}\right)\,.
\end{equation}

In the high-curvature limit, we have
\begin{equation}
    F(R\gg R_{\mathrm{c}})=R-R_{\mathrm{c}}\ln\left(1+b\right)\,.
\end{equation}
To mimic the $\Lambda$CDM model in the high-curvature limit, we require that $\lambda R_{\mathrm{c}}=2\Lambda$, with $\lambda$ defined by
\begin{equation}
    \lambda \equiv\ln\left(1+b\right)\,.
\end{equation}
The resulting $F(R)$ model is then
\begin{equation}
    \label{eq:logistic_model}
    F(R)=R-\lambda R_{\mathrm{c}}+R_{\mathrm{c}}\ln\left[1+\left(e^{\lambda}-1\right)e^{-R/R_{\mathrm{c}}}\right]
    \,.
\end{equation}

To quantify how it deviates from the $\Lambda$CDM model, we further parameterize the model as
\begin{equation}
    \label{eq:parameterized_logistic_model}
    F(R)=R-2\Lambda+\xi\Lambda\ln\left[1+\left(e^{2/\xi}-1\right)e^{-R/\xi\Lambda}\right]
    \,,
\end{equation}
where $\xi\equiv 2/\lambda=R_{\mathrm{c}}/\Lambda$ characterizes the relative size of $R_\mathrm{c}$ with respect to $\Lambda$.
One finds that the parameter $\xi$ controls the steepness of the curve: the smaller $\xi$ is, the steeper the function becomes, and the closer the $F(R)$ model is to the $\Lambda$CDM model.
For $\xi = 0$, $F_R$ reduces to a Heaviside step function at $R = 0$, and the model recovers the $\Lambda$CDM limit (for $R > 0$).
For fixed curvature, the model can be approximated as follows:
\begin{equation}
    F\left(R|R>2\Lambda\right)=
    \begin{cases}
        R-2\Lambda\,, & \xi\to +0\,,     \\
        R             & \xi\to+\infty\,.
    \end{cases}
\end{equation}

The new form \eqref{eq:logistic_model} provides an alternative understanding of how each parameter and each term works.
Furthermore, one can observe how it differs from the Tsujikawa and exponential models, which also contain exponential terms.
The key point of Eq.~\eqref{eq:logistic_model} to guarantee $F_{R}>0$ and $F_{RR}>0$ for $R<0$ is that the nonlinear term in Eq.~\eqref{eq:logistic_model} is indeed "linearized" when $R\ll -R_\mathrm{c}$;  this results in an absolute Ricci scalar $|R|$, which is canceled by the linear term $R$.

The generalization of model~\eqref{eq:logistic_fr} is straightforward:
\begin{equation}
    \label{eq:logistic_function_n}
    F_{R}(R)=\frac{1}{\left(1+be^{-R/R_{\mathrm{c}}}\right)^{n}}\,,
\end{equation}
where $b,n>0$.
It is positive for all $R$, and its derivative is also positive,
\begin{equation}
    F_{RR}(R) =\frac{nb}{R_{\mathrm{c}}}\frac{e^{-R/R_{\mathrm{c}}}}{\left(1+be^{-R/R_{\mathrm{c}}}\right)^{n+1}}>0\,.
\end{equation}
One observes that $F_{R}(R)$ and $F_{R}(RR)$ decrease faster as $n$ increases.

Integrating Eq.~\eqref{eq:logistic_function_n} yields
\begin{equation}
    F(R)=\frac{R_{\mathrm{c}}}{nb}\left(b+e^{R/R_{\mathrm{c}}}\right)\left(be^{-R/R_{\mathrm{c}}}+1\right)^{-n}\,_{2}F_{1}\left(1,1;n+1;-e^{R/R_{\mathrm{c}}}/b\right)+C\,,
\end{equation}
where $C$ is an integration constant and $\,_{2}F_{1}\left(1,1;n+1;-\frac{e^{x}}{b}\right)$ is a hypergeometric function.
It has a simple form if $n\in \mathbb{Z}$; for instance, when $n=2$, we have
\begin{equation}
    F(R;n=2) =R-\ln\left(1+b\right)R_{\mathrm{c}}-\frac{1-e^{-R/R_{\mathrm{c}}}}{1+be^{-R/R_{\mathrm{c}}}}\frac{b}{1+b}R_{\mathrm{c}}+R_{\mathrm{c}}\ln\left(1+be^{-R/R_{\mathrm{c}}}\right)\,,
\end{equation}
where the condition $F(0)=0$ has been applied to eliminate the integration constant.
Further exploration of the generalized model is left for future work.

\subsection{New model}

Having shown the benefit of the sigmoid $F_{R}$ function and reproduced the original Appleby-Battye model, we now consider models that exhibit a power-law behavior in the high-curvature limit.
A simple sigmoid $F_{R}$ function is
\begin{equation}
    F_{R}(R)=\frac{\sqrt{\left(R/R_{\mathrm{c}}-b\right)^{2}+c}+R/R_{\mathrm{c}}-b}{2\sqrt{\left(R/R_{\mathrm{c}}-b\right)^{2}+c}}\,,
\end{equation}
where $b,c$ are free parameters and $R_{\mathrm{c}}$ is the critical curvature.
It is required that $c>0$ for model stability, and we impose that $R_{\mathrm{c}}>0$ with $b$ unconstrained.

The midpoint of the function is $\left(bR_{\mathrm{c}}, \frac{1}{2}\right)$.
For fixed curvature $R_{\mathrm{c}}$, $b$ controls the shift of the midpoint and $c$ controls the clockwise rotation:
the smaller $c$ is, the steeper the curve becomes, and the closer the model is to the $\Lambda$CDM model.
When $c=0$, it becomes a Heaviside delta function:
\begin{equation}
    F_{R}\left(R|c=0\right)=\frac{\left|R/R_{\mathrm{c}}-b\right|+R/R_{\mathrm{c}}-b}{2\left|R/R_{\mathrm{c}}-b\right|}=\begin{cases}
        1\,, & R/R_{\mathrm{c}}>b,           \\
        1\,, & R/R_{\mathrm{c}}-b\to0^{+}\,, \\
        0\,, & R/R_{\mathrm{c}}-b\to0^{-}    \\
        0\,, & R<bR_{\mathrm{c}}\,.
    \end{cases}
\end{equation}
Therefore, The $F(R)$ model is indistinguishable from the $\Lambda$CDM model when $c=0$ and $R>bR_{\mathrm{c}}$.

Integrating $F_{R}(R)$ yields the primitive function
\begin{equation}
    \label{eq:sqrt_f(R)}
    F(R)=\frac{R}{2}+\frac{R_{\mathrm{c}}}{2}\sqrt{\left(R/R_{\mathrm{c}}-b\right)^{2}+c}+C\,,
\end{equation}
where $C$ is an integration constant, which can be determined by imposing $F(0)=0$:
\begin{equation}
    C=-\frac{R_{\mathrm{c}}}{2}\sqrt{b^{2}+c}\,.
\end{equation}
Then, the full $F(R)$ function takes the form
\begin{equation}
    \label{eq:sqrt_model}
    F(R)=\frac{R}{2}+\frac{R_{\mathrm{c}}}{2}\sqrt{\left(R/R_{\mathrm{c}}-b\right)^{2}+c}-\frac{R_{\mathrm{c}}}{2}\sqrt{b^{2}+c}\,.
\end{equation}

In the high-curvature limit,
\begin{equation}
    F\left(R\gg R_{\mathrm{c}}\right)\approx R-\frac{R_{\mathrm{c}}}{2}\left(b+\sqrt{b^{2}+c}\right)+\frac{cR_{\mathrm{c}}}{4}\frac{1}{R/R_{\mathrm{c}}-b}\,.
\end{equation}
To mimic the $\Lambda$CDM model, we impose $\lambda R_{\mathrm{c}}=2\Lambda$, with $\lambda$ defined by
\begin{equation}
    \lambda\equiv\frac{1}{2}\left(b+\sqrt{b^{2}+c}\right)\,,\label{eq:sqrt_para_lambda}
\end{equation}
where $c,\lambda>0$, and the parameter $b$ is constrained to be
\begin{equation}
    b<\lambda\,.
\end{equation}
Therefore, the alternative form is
\begin{equation}
    \label{eq:sqrt_lambda}
    F(R)=R-\lambda R_{\mathrm{c}}+\frac{R_{\mathrm{c}}}{2}\sqrt{\left(R/R_{\mathrm{c}}-b\right)^{2}+4\lambda\left(\lambda-b\right)}-\frac{R_{\mathrm{c}}}{2}\left(R/R_{\mathrm{c}}-b\right)\,.
\end{equation}
One observes that, when $R/R_{\mathrm{c}}\gg b$, the above equation reduces to $F(R)=R-2\Lambda$.
On the other hand, the relation $R>\lambda R_{\mathrm{c}}>bR_{\mathrm{c}}$ yields the following asymptotic behaviors:
\begin{equation}
    F\left(R\right)=\begin{cases}
        R-2\Lambda\,, & b\to\lambda\,, \\
        R\,,          & b\to-\infty\,.
    \end{cases}
\end{equation}
It can be observed that $b$ can well describe the deviation of this model with respect to the $\Lambda$CDM model.
Therefore, we can parameterize the model by defining a dimensionless parameter
\begin{equation}
    \xi\equiv1-\frac{b}{\lambda}=\frac{c}{4\lambda^{2}}\,.\label{eq:sqrt_xi}
\end{equation}
Then, the model becomes
\begin{equation}
    \label{eq:parameterized_sqrt}
    F(R)=R-2\Lambda+\Lambda\sqrt{\left(R/2\Lambda+\xi-1\right)^{2}+4\xi}-\Lambda\left(R/2\Lambda+\xi-1\right)\,.
\end{equation}
It has the following asymptotic behaviors:
\begin{equation}
    F\left(R|R>2\Lambda\right)=\begin{cases}
        R-2\Lambda\,,                                                & \xi\to0^{+}\,,   \\
        R-2\Lambda+2\Lambda\frac{\xi}{\left(R/2\Lambda+\xi-1\right)} & \xi>1\,,         \\
        R\,,                                                         & \xi\to+\infty\,.
    \end{cases}
\end{equation}
The allowed parameter space of $\xi$ is $0<\xi<+\infty$, whereas observational constraints suggest $\xi\ll1$.
Furthermore, we observe that $b$ and $c$ are not entirely independent according to the definition of $\xi$.
Following Ref. \cite{Hu:2007nk}, one can fix $R_{\mathrm{c}}\equiv\frac{8\pi G}{3}\rho_{\mathrm{m}}$,
such that the physical density parameter inferred from the $\Lambda$CDM model applies to $F(R)$ models as well.
With $R_\mathrm{c}$ determined, once $\xi$ is known from the observation, other parameters are also determined, $\xi\rightarrow b\rightarrow\lambda\rightarrow c$.
Therefore, our model has only one additional free parameter compared with the $\Lambda$CDM model.
Moreover, $\xi\ll1$ indicates that $b\to\lambda$ and $c\ll4\lambda^{2}$.

Finally, the generalization of this model to higher powers is straightforward:
\begin{equation}
    F(R)=\frac{R}{2}+\frac{R_{\mathrm{c}}}{2}\left[\left(R/R_{\mathrm{c}}-b\right)^{2n}+c\right]^{1/2n}+C\,.
\end{equation}
where $C$ is a constant. Imposing $F(0)=0$, we have
\begin{equation}
    C=-\frac{R_{\mathrm{c}}}{2}\left(b^{2n}+c\right)^{1/2n}\,.
\end{equation}
Then, the full function takes the form
\begin{equation}
    F(R)=\frac{R}{2}+\frac{R_{\mathrm{c}}}{2}\left[\left(R/R_{\mathrm{c}}-b\right)^{2n}+c\right]^{1/2n}-\frac{R_{\mathrm{c}}}{2}\left(b^{2n}+c\right)^{1/2n}\,.
\end{equation}
In the high-curvature limit,
\begin{equation}
    F(R\gg R_{\mathrm{c}})\approx R-\frac{R_{\mathrm{c}}}{2}\left[b+\left(b^{2n}+c\right)^{1/2n}\right]+\frac{cR_{\mathrm{c}}}{4n\left(R/R_{\mathrm{c}}-b\right)^{2n-1}}
\end{equation}
We require $\lambda R=2\Lambda$, with $\lambda$ defined by
\begin{equation}
    \label{eq:power-law_lambda}
    \lambda\equiv\frac{1}{2}\left[b+\left(b^{2n}+c\right)^{1/2n}\right]\,.
\end{equation}
Here, $c,\lambda>0$, and the parameter $b$ is confined to
\begin{equation}
    b<\lambda\,.
\end{equation}
Therefore, the alternative $F(R)$ form reads
\begin{equation}
    \label{eq:power-law}
    F(R)=R-\lambda R_{\mathrm{c}}+\frac{R_{\mathrm{c}}}{2}\left[\left(R/R_{\mathrm{c}}-b\right)^{2n}+\left(2\lambda-b\right)^{2n}-b^{2n}\right]^{1/2n}-\frac{R_{\mathrm{c}}}{2}\left(R/R_{\mathrm{c}}-b\right)\,.
\end{equation}

Similarly, by defining $\xi\equiv1-\frac{b}{\lambda}$, the model can be parameterized as
\begin{equation}
    F(R)=R-2\Lambda+\Lambda\left[\left(R/2\Lambda+\xi-1\right)^{2n}+\left(1+\xi\right)^{2n}-\left(1-\xi\right)^{2n}\right]^{1/2n}-\Lambda\left(R/2\Lambda+\xi-1\right)\,.
\end{equation}
For later convenience, we define $x\equiv R/\Lambda$ and rewrite the model as the following dimensionless forms:
\begin{equation}
    \label{eq:normalized_power-law}
    \begin{aligned}F(R)/\Lambda   & =x-2+\left[\left(x/2+\xi-1\right)^{2n}+\left(1+\xi\right)^{2n}-\left(1-\xi\right)^{2n}\right]^{1/2n}-\left(x/2+\xi-1\right)\,,                                                                                                 \\
               F_{R}(R)       & =\frac{1}{2}+\frac{1}{2}\frac{\left(x/2+\xi-1\right)^{2n-1}}{\left[\left(x/2+\xi-1\right)^{2n}+\left(1+\xi\right)^{2n}-\left(1-\xi\right)^{2n}\right]^{\frac{2n-1}{2n}}}\,,                                                    \\
               \Lambda F_{RR} & =\frac{2n-1}{4}\frac{\left(x/2+\xi-1\right)^{2n-2}\left[\left(1+\xi\right)^{2n}-\left(1-\xi\right)^{2n}\right]}{\left[\left(x/2+\xi-1\right)^{2n}+\left(1+\xi\right)^{2n}-\left(1-\xi\right)^{2n}\right]^{\frac{4n-1}{2n}}}\,.
    \end{aligned}
\end{equation}
One can readily observe that $F_{RR}>0$ if $n>0.5$, and the $F_{R}(R)$ function satisfies
\begin{equation}
    F_{R}(R)=
    \begin{cases}
        1\,, & R\to+\infty\,, \\
        0\,, & R\to-\infty\,.
    \end{cases}
\end{equation}

In the high-curvature limit, Eqs.~\eqref{eq:normalized_power-law} can be approximated to (up to the first order of $\xi$)
\begin{equation}
    \begin{aligned}
        f/\Lambda
         & \approx-2+\left(\frac{2}{x-2}\right)^{2n-1}2\xi\,,                \\
        f_{R}
         & \approx-\left(2n-1\right)\left(\frac{2}{x-2}\right)^{2n}\xi\,,    \\
        \Lambda f_{RR}
         & \approx n\left(2n-1\right)\left(\frac{2}{x-2}\right)^{2n+1}\xi\,.
    \end{aligned}
    \label{eq:power-law_sim}
\end{equation}

\begin{figure}
    \centering
    \includegraphics[width=\textwidth]{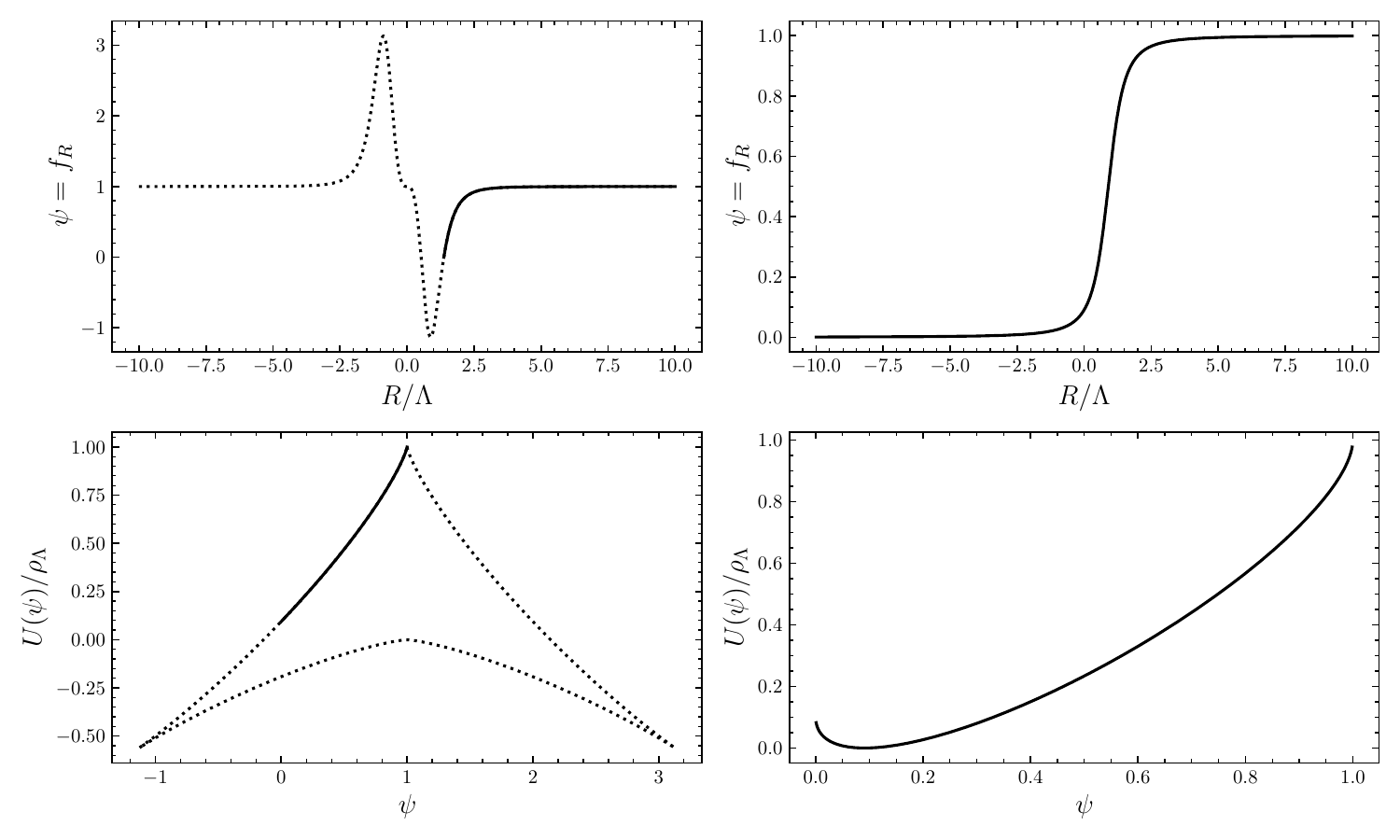}
    \caption{Comparison of the Jordan-frame field (upper panel) and potential (lower panel) between the Hu-Sawicki model (left panel with $n=1$ and $\xi=0.1$) and our model (right panel with $n=1$ and $\xi=0.1$). Here, $\rho_\Lambda=\frac{\Lambda}{\kappa^2}$.}
    \label{fig:fR_U_cmp}
\end{figure}

Note that the $\xi$s in the two classes of models have different meanings.
In the Appleby-Battye model, $\xi\equiv\frac{2}{\lambda}=\frac{R_{\mathrm{c}}}{\Lambda}$, whiche measures the relative size of the critical curvature with respect to the cosmological constant, whereas in our model,
\begin{equation}
    \xi \equiv\frac{2\Lambda-bR_{\mathrm{c}}}{2\Lambda}\,,
\end{equation}
which measures the relative difference of $b R_{\mathrm{c}}$ with respect to $2\Lambda$.
Therefore, to satisfy the observations, $\xi\ll 1$ in our model, whereas $\xi<1$ in the Appleby-Battye model.
It should be noted that the potential for most of $F(R)$ dark energy models is multi-valued, and only the branch corresponding to $F_{R}>0$ is physical.
Using the Hu-Sawicki model as an example, as illustrated in the left panel of Fig.~\ref{fig:fR_U_cmp}, only the black sections of the curves are physically relevant.
This is not true in globally stable models as $F_{R}>0$ for all $R$.
As shown in Fig.~\ref{fig:fR_U_cmp}, the multi-valued nature of the potential naturally disappears in our model.
Therefore, the scalar fields defined in both frames exhibit a one-to-one correspondence.

\section{Cosmological evolution}

We now study the cosmological evolution by explicitly solving the Friedmann equation.
In the flat Friedmann-Lema\^{i}tre-Robertson-Walker spacetime, the modified Friedmann equation is
\begin{equation}
    H^{2}=\frac{8\pi G}{3}\rho_{\mathrm{tot}}=\frac{8\pi G}{3}\left(\rho+\rho_{\mathrm{de}}\right)\,,
\end{equation}
where we have defined the effective energy density rising from modified gravity as
\begin{equation}
    \rho_{\mathrm{de}}\equiv\frac{3}{8\pi G}\left[\frac{1}{6}\left(Rf_{R}-f\right)-H\dot{f_{R}}-H^{2}f_{R}\right]\,,
\end{equation}
Analogously, the modified acceleration equation is
\begin{equation}
    2\dot{H}+3H^{2} =-8\pi GP_{\mathrm{tot}}=-8\pi G\left(P+P_{\mathrm{de}}\right)\,,
\end{equation}
where the effective pressure is
\begin{equation}
    P_{\mathrm{de}}=\frac{1}{8\pi G}\left[\ddot{f_{R}}+2H\dot{f_{R}}+\left(2\dot{H}+3H^{2}\right)f_{R}-\frac{1}{2}\left(Rf_{R}-f\right)\right]\,.
\end{equation}
Hence, the effective EoSs of the system and dark energy are defined by
\begin{equation}
    w_{\mathrm{eff}} \equiv\frac{P_{\mathrm{tot}}}{\rho_{\mathrm{tot}}}=-1-\frac{2\dot{H}}{3H^{2}}\,,
\end{equation}
and
\begin{equation}
    w_{\mathrm{de}}
    \equiv\frac{P_{\mathrm{de}}}{\rho_{\mathrm{de}}}
    =\frac{3H^{2}w_{\mathrm{eff}}-8\pi GP}{3H^{2}-8\pi G\rho}\,,
\end{equation}
respectively.
Substituting $\rho\propto a^{-3(1+w)}$ into the Friedmann equation gives
\begin{equation}
    H^{2}+H^{2}f_{R}-\frac{1}{6}\left(Rf_{R}-f\right)+Hf_{RR}\dot{R}=H_{0}^{2}\left(\Omega_{\mathrm{m}0}\left(1+z\right)^{3}+\Omega_{\mathrm{r}0}\left(1+z\right)^{4}\right)\,.
\end{equation}
where the density parameter $\Omega_{i}$ and Hubble parameter $H_{0}$ are model-dependent, i.e.,
\begin{equation}
    \Omega_{i0} \neq\Omega_{i0}^{\Lambda\mathrm{CDM}}\,,\qquad H_{0}\neq H_{0}^{\Lambda\mathrm{CDM}}\,,
\end{equation}
However, their combination, the physical density parameter $\omega_{i}\equiv\Omega_{i}h^{2}$, is usually model-independent.
Therefore,
\begin{equation}
    \Omega_{i0}H_{0}^{2}=\Omega_{i0}^{\Lambda\mathrm{CDM}}\left(H_{0}^{\Lambda\mathrm{CDM}}\right)^{2}\,.
\end{equation}
This relation is convenient when making comparisons.

We now rewrite the equations in a dimensionless form.
We first define the dimensionless Hubble parameter $E\equiv H/H_{0}^{\Lambda\mathrm{CDM}}$ (in the following, we neglect the notation $\Lambda$CDM) and replace the cosmic time $t$ by redshift $z$.
Then, the Friedmann equation becomes
\begin{equation}
    \begin{aligned}
        \frac{2}{\Omega_{\Lambda0}}E^{3}\left(1+z\right)^{2}\left(E_{zz}+\frac{E_{z}^{2}}{E}-\frac{3E_{z}}{1+z}\right)\Lambda f_{RR}+\left[\left(1+z\right)EE_{z}-E^{2}\right]f_{R} \\
        +\left[E^{2}+\frac{\Omega_{\Lambda0}}{2}\frac{f}{\Lambda}-\Omega_{\mathrm{m}0}\left(1+z\right)^{3}-\Omega_{\mathrm{r}0}\left(1+z\right)^{4}\right] & =0\,,
    \end{aligned}
    \label{eq:normalized_friedmann}
\end{equation}
where the subscript "$z$" denotes the derivative with respect to $z$.
According to $f(R)$ and its derivatives, the above three terms, represent the second-order, first-order, and zeroth-order corrections of the Friedmann equation in $F(R)$ gravity with respect to the $\Lambda$CDM model (with $f=-2\Lambda$), respectively.
In addition, we have written all quantities in dimensionless forms, i.e., $f(R)/\Lambda$, $f_{R}(R)$ and $\Lambda f_{RR}$.
The Ricci scalar is normalized by
\begin{equation}
    R/\Lambda =\frac{2E}{\Omega_{\Lambda0}}\left(2E-\left(1+z\right)E_{z}\right)\,,
\end{equation}
where $\Omega_{\Lambda0}=\Lambda/3H_{0}^{2}$.

We assume that the $F(R)$ model reduces to the $\Lambda$CDM model in the high redshift to solve the modified Friedmann equation.
Then, the initial conditions can be inferred from the $\Lambda$CDM model:
\begin{equation}
    \begin{aligned}E     & =\sqrt{\Omega_{\mathrm{r0}}\left(1+z\right)^{4}+\Omega_{\mathrm{m0}}\left(1+z\right)^{3}+\Omega_{\Lambda0}}\,,   \\
               E_{z} & =\frac{1}{2E}\left[4\Omega_{\mathrm{r0}}\left(1+z\right)^{3}+3\Omega_{\mathrm{m0}}\left(1+z\right)^{2}\right]\,.
    \end{aligned}
\end{equation}
Therefore,
\begin{equation}
    \begin{aligned}E(z_{\mathrm{ini}})     & = E^{\Lambda\mathrm{CDM}}(z_{\mathrm{ini}})\,,     \\
               E_{z}(z_{\mathrm{ini}}) & = E_{z}^{\Lambda\mathrm{CDM}}(z_{\mathrm{ini}})\,.
    \end{aligned}
\end{equation}
After solving $E(z)$ and $E_{z}(z)$, the calculation of the eff EoSs for the system and dark energy component is straightforward:
\begin{align}
    w_{\mathrm{eff}} & =-1+\frac{2E_{z}\left(1+z\right)}{3E}\,,                                                                                                                \\
    w_{\mathrm{de}}  & =-1+\frac{2}{3}\left(1+z\right)\frac{EE_{z}-E^{\Lambda\mathrm{CDM}}E_{z}^{\Lambda\mathrm{CDM}}}{E^{2}-E_{\Lambda\mathrm{CDM}}^{2}+\Omega_{\Lambda0}}\,.
\end{align}
Note that the density parameters used here are those from the $\Lambda$CDM model.
The resulting Hubble constant and density parameter are
\begin{align}
    H_{0}                & =H_{0}^{\Lambda\mathrm{CDM}}E(z=0)\,,                             \\
    \Omega_{\mathrm{m}0} & =\frac{\Omega_{\mathrm{m}0}^{\Lambda\mathrm{CDM}}}{E^{2}(z=0)}\,.
\end{align}

\begin{figure}
    \centering
    \includegraphics[scale=0.6]{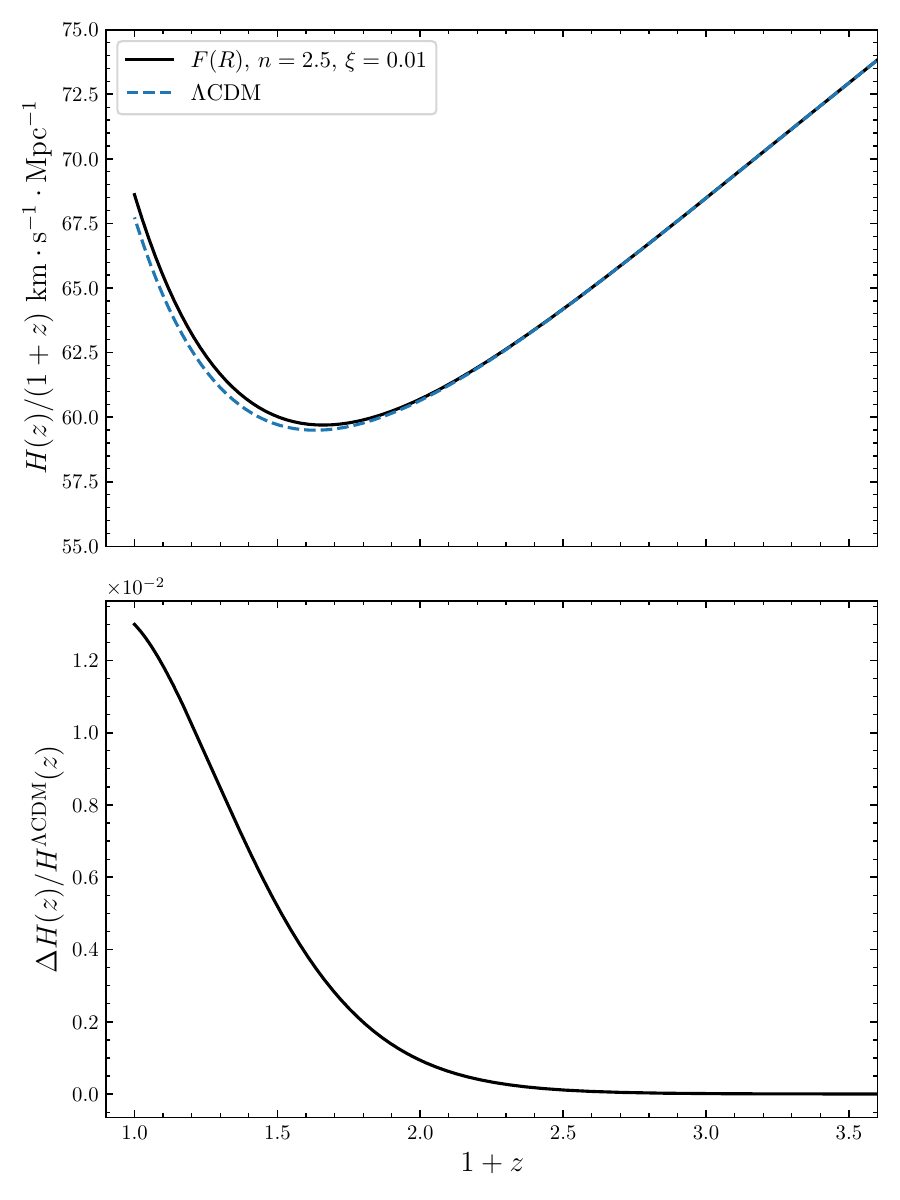}
    \caption{Hubble parameter $H$ as a function of $1+z$, with $n=2.5$ and $\xi=0.01$. The lower panel shows the relative difference $\frac{H(z)-H^{\Lambda\mathrm{CDM}}(z)}{H^{\Lambda\mathrm{CDM}}(z)}$.}
    \label{fig:H_z}
\end{figure}

\begin{figure}
    \centering
    \includegraphics[scale=0.6]{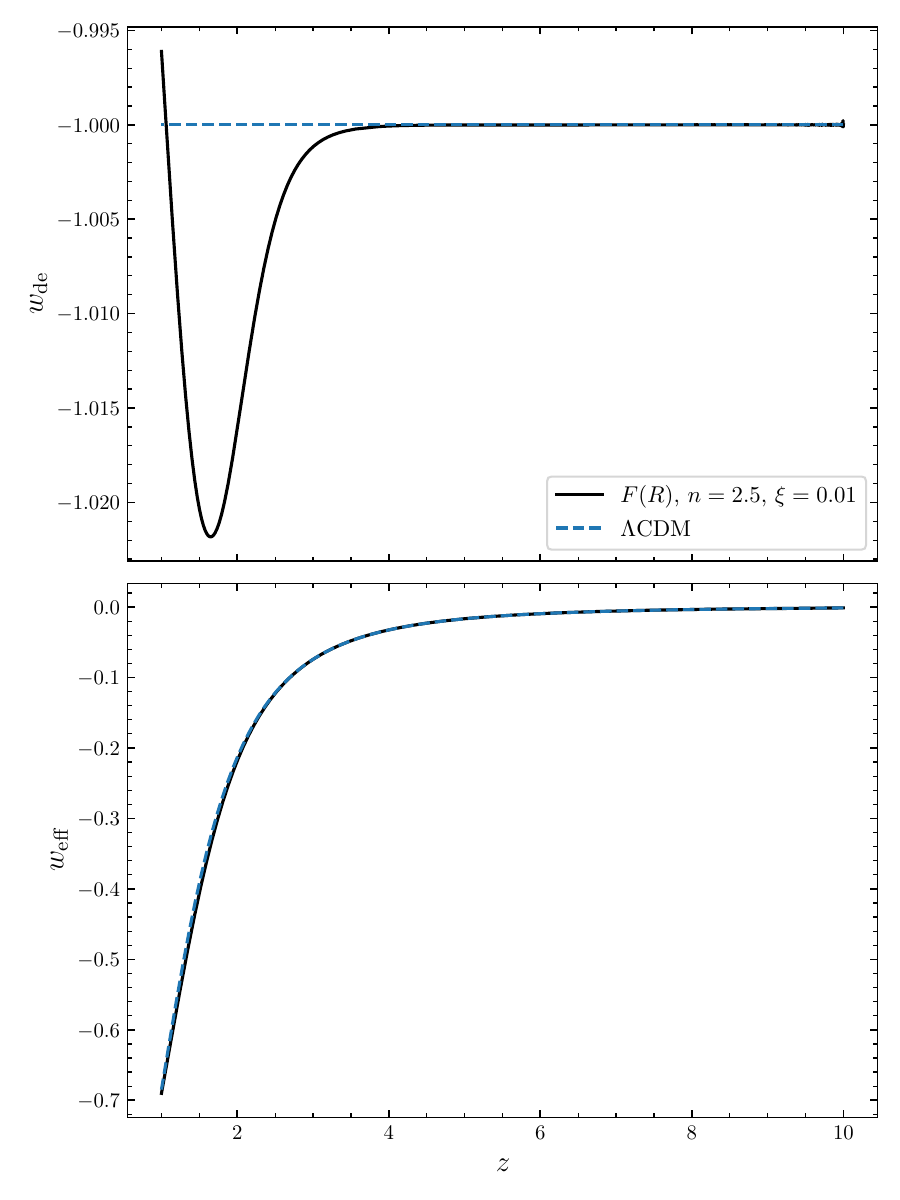}
    \caption{Comparison of the EoS parameter between our model and the $\Lambda$CDM model, with $n=2.5$ and $\xi=0.01$.
        The upper and lower panels show the effective EoS of dark energy, $w_{\mathrm{de}}$, and the effective EoS of the entire system, $w_{\mathrm{eff}}$, respectively.}
    \label{fig:wde_weff}
\end{figure}

Substituting our model into Eq.~\eqref{eq:normalized_friedmann}, we can extract the Hubble parameter and EoS.
As shown in Fig.~\ref{fig:H_z}, the Hubble parameter in $F(R)$ deviates from that in the $\Lambda$CDM model when $z<2$, resulting in a slightly larger Hubble constant.
Figure~\ref{fig:wde_weff} compares the differences in EoS between the two models.
We observe that the two models exhibit a similar evolution of the EoS of the system; however, the EoS of the dark energy component shows significant differences: there is a phantom crossing in our model when $z<4$ (with the initial values chosen above).
This behavior is a universal characteristic of $F(R)$ dark energy models \cite{Hu:2007nk,Bamba:2010iy,Motohashi:2010zz,Motohashi:2010tb,Motohashi:2011wy,Jaime:2013zwa}.
Therefore, our model can be distinguished from the $\Lambda$CDM model.

\section{Local gravity tests}

In this section, we confront our model with local gravity tests.
For the calculation details, refer to Refs. \cite{Khoury:2003aq, Khoury:2003rn, Faulkner:2006ub, Capozziello:2007eu, DeFelice:2010aj}.
For the latest reviews on chameleons, see \cite{Burrage:2017qrf, Fischer:2024eic}.

\subsection{Chameleon mechanism}

In the Einstein frame, the scalaron couples to the trace of energy-momentum tensor matter with a universal coupling constant $\beta=-1/\sqrt{6}$.
In the dense region, we have $T_{\mu}^{\mu}\approx-\rho_{\mathrm{m}}$.
We consider a spherical static body with radius $\tilde{r}_{\mathrm{c}}$, in which case the background spacetime is approximated to the Minkowski one.
The variation with respect to Eq.~\eqref{eq:ef_action} yields
\begin{equation}
    \frac{d^{2}\varphi}{d\tilde{r}^{2}}+\frac{2}{\tilde{r}}\frac{d\varphi}{d\tilde{r}}-\frac{dV_{\mathrm{eff}}}{d\varphi}=0\,,\label{eq:sss}
\end{equation}
where the effective potential is
\begin{equation}
    V_{\mathrm{eff}}(\varphi)=V(\varphi)+e^{\beta\kappa\varphi}\tilde{\rho}_{*}\,.
\end{equation}
$\tilde{\rho}_{*}$ is the conserved energy density in the Einstein frame \eqref{eq:ef_conserved_density} satisfying $\tilde{r}^{3}\tilde{\rho}_{*}=r^{3}\rho$.
For a given mass density $\tilde{\rho}_{*}$, the effective potential minimum is determined by
\begin{equation}
    V_{,\varphi}(\varphi_{\mathrm{min}})+\beta\kappa e^{\beta\kappa\varphi_{\mathrm{min}}}\tilde{\rho}_{*}=0\,.
\end{equation}
In dense environments, such as our galaxy and solar system, $R\gg R_{\mathrm{c}}$; hence we can use the approximated forms of our model \eqref{eq:power-law_sim}.
Then, the potential is approximated as
\begin{equation}
    V(\varphi)\approx\rho_{\Lambda}e^{4\beta\kappa\varphi}\left[1-2\beta\kappa\varphi-2n\left[\frac{2\beta\kappa\varphi}{\left(2n-1\right)\xi}\right]^{\frac{2n-1}{2n}}\xi\right]\,,
\end{equation}
where $\rho_{\Lambda}=\frac{\Lambda}{\kappa^{2}}$.

For convenience, we assume that the density distribution is constant both inside ($\tilde{r}<\tilde{r}_{\mathrm{c}}$) and outside ($\tilde{r}>\tilde{r}_{\mathrm{c}}$) the body, denoted by $\tilde{\rho}_{\mathrm{in}}$ and $\tilde{\rho}_{\mathrm{out}}$, respectively.
The densities satisfy $\tilde{\rho}_{\mathrm{out}}\ll\tilde{\rho}_{\mathrm{in}}$.
Then, the mass and gravitational potential are
\begin{align}
    M_{\mathrm{c}}    & =\frac{4\pi}{3}\tilde{r}_{\mathrm{c}}^{3}\tilde{\rho}_{\mathrm{in}}\,,                                                      \\
    \Phi_{\mathrm{c}} & =\frac{GM_{\mathrm{c}}}{\tilde{r}_{\mathrm{c}}}=\frac{\tilde{r}_{\mathrm{c}}^{2}}{6}\kappa^{2}\tilde{\rho}_{\mathrm{in}}\,.
\end{align}
$R\gg R_{\mathrm{c}}$; therefore, we have $R\simeq\tilde{\rho}_{*}$, and
\begin{equation}
    \varphi\approx\frac{2n-1}{2\kappa\beta}\left(\frac{2\rho_{\Lambda}}{\tilde{\rho}_{*}}\right)^{2n}\xi\,.
    \label{eq:power-law_phi_sss}
\end{equation}
$\rho_{\Lambda}\ll\tilde{\rho}_{\mathrm{out}}\ll\tilde{\rho}_{\mathrm{in}}$ indicates $\left|\kappa\varphi_{\mathrm{in}}\right|\ll\left|\kappa\varphi_{\mathrm{out}}\right|\ll1$.
The effective potential has two minima at $\varphi_{\mathrm{in}}$ and $\varphi_{\mathrm{out}}$ corresponding to $V_\mathrm{eff}^{\prime}(\varphi_{\mathrm{in}})=0$ and $V_\mathrm{eff}^{\prime}(\varphi_{\mathrm{out}})=0$.
As $\rho_{\Lambda}\ll\tilde{\rho}_{\mathrm{out}}\ll\tilde{\rho}_{\mathrm{in}}$, we have $\left|\kappa\varphi_{\mathrm{in}}\right|\ll\left|\kappa\varphi_{\mathrm{out}}\right|\ll1$.
The effective mass of the scalaron is
\begin{equation}
    m_{\varphi}^{2}\equiv V_{\mathrm{eff},\varphi\varphi}(\varphi_{\min})\approx\frac{2\beta^{2}\Lambda}{n\left(2n-1\right)\xi}\left(\frac{\tilde{\rho}_{*}}{2\rho_{\Lambda}}\right)^{2n+1}\,.
\end{equation}

Eq.~\eqref{eq:sss} can be regarded as an equation of motion in terms of time $\tilde{r}$ under the effective force term $-\partial V_{\mathrm{eff}}/\partial\varphi$ and friction term $\frac{2}{\tilde{r}}\frac{d\varphi}{d\tilde{r}}$.
We assume the following boundary conditions:
\begin{equation}
    \begin{aligned}\frac{d\varphi}{d\tilde{r}}\left(\tilde{r}=0\right) & =0\,,                      \\
               \varphi\left(\tilde{r}\to\infty\right)              & =\varphi_{\mathrm{out}}\,.
    \end{aligned}
    \label{eq:boundary_conditions}
\end{equation}
By solving Eq.~\eqref{eq:sss} the field profile outside the body is approximated to
\begin{equation}
    \label{eq:outter_field_profile}
    \varphi\left(\tilde{r}\right)\approx\varphi_{\mathrm{out}}-\frac{2\beta_{\mathrm{eff}}}{\kappa}\frac{GM_{\mathrm{c}}}{\tilde{r}}e^{-m_{\mathrm{out}}\left(\tilde{r}-\tilde{r}_{\mathrm{c}}\right)}\,,
\end{equation}
where the effective coupling $\beta_{\mathrm{eff}}\equiv3\beta\epsilon_{\mathrm{th}}$ is defined in terms of the thin-shell parameter
\begin{equation}
    \label{eq:thin-shell_parameter}
    \epsilon_{\mathrm{th}}\equiv\frac{\kappa\left(\varphi_{\mathrm{out}}-\varphi_{\mathrm{in}}\right)}{6\beta\Phi_{\mathrm{c}}}\,.
\end{equation}
The condition $\epsilon_{\mathrm{th}}\ll 1$ corresponds to the so-called thin-shell effect, significantly reducing the effective coupling.

\subsection{Solar system constraints}

We shall first confront our model with the post-Newtonian solar system tests.
The spherical symmetric metric in the Einstein frame is
\begin{equation}
    d\tilde{s}^{2}=\tilde{g}_{\mu\nu}\mathrm{d}\tilde{x}^{\mu}\mathrm{d}\tilde{x}^{\nu}=-\left(1-2\tilde{\mathcal{A}}(\tilde{r})\right)\mathrm{d}t^{2}+\left(1+2\tilde{\mathcal{B}}(\tilde{r})\right)\mathrm{d}\tilde{r}^{2}+\tilde{r}^{2}\mathrm{d}\Omega^{2}\,.
\end{equation}
In the weak-field approximation, we have $\tilde{\mathcal{A}}(\tilde{r})\ll1$ and $\tilde{\mathcal{B}}(\tilde{r})\ll1$.
In addition, $\tilde{\mathcal{A}}(\tilde{r})\simeq\tilde{\mathcal{B}}(\tilde{r})\simeq\frac{GM_{\mathrm{c}}}{\tilde{r}}$ outside the body.
The Jordan-frame metric is written as
\begin{equation}
    ds^{2}=g_{\mu\nu}\mathrm{d}x^{\mu}\mathrm{d}x^{\nu}=-\left(1-2\mathcal{A}(r)\right)\mathrm{d}t^{2}+\left(1+2\mathcal{B}(r)\right)\mathrm{d}r^{2}+r^{2}\mathrm{d}\Omega^{2}\,.
\end{equation}
For $e^{2\beta\kappa\varphi}\simeq1+2\beta\kappa\varphi$, the gravitational potentials transform as
\begin{align}
    \mathcal{A}(r) & =\tilde{\mathcal{A}}(\tilde{r})-\beta\kappa\varphi(\tilde{r})\,,             \\
    \mathcal{B}(r) & =\tilde{\mathcal{B}}(\tilde{r})-\beta\kappa\frac{d\varphi}{d\ln\tilde{r}}\,.
\end{align}
$\left|\varphi_{\mathrm{out}}\right|\gg\left|\varphi_{\mathrm{in}}\right|$ implies that
\begin{equation}
    \varphi_{\mathrm{out}}\approx\frac{6\beta\epsilon_{\mathrm{th}}}{\kappa}\frac{GM_{\mathrm{c}}}{\tilde{r}_{\mathrm{c}}}=\frac{6\beta\epsilon_{\mathrm{th}}\Phi_{\mathrm{c}}}{\kappa}\,.
\end{equation}
Setting $\tilde{r}\simeq r$, the thin-shell solution \eqref{eq:outter_field_profile} at $r\gtrsim r_{\mathrm{c}}$ can be approximated to
\begin{equation}
    \varphi(r) \approx\frac{6\beta\epsilon_{\mathrm{th}}}{\kappa}\frac{GM_{\mathrm{c}}}{r}\left[\frac{r}{r_{\mathrm{c}}}-1\right]\,.\label{eq:filed_profile}
\end{equation}
Substituting the above into the potentials gives
\begin{align}
    \mathcal{A}(r) & =\frac{GM_{\mathrm{c}}}{r}\left[1+6\beta^{2}\epsilon_{\mathrm{th}}\left(1-\frac{r}{r_{\mathrm{c}}}\right)\right]\,, \\
    \mathcal{B}(r) & =\frac{GM_{\mathrm{c}}}{r}\left(1-6\beta^{2}\epsilon_{\mathrm{th}}\right)\,.
\end{align}
Therefore, the PPN parameter is
\begin{equation}
    \gamma\equiv\frac{\mathcal{B}(r)}{\mathcal{A}(r)}\simeq\frac{1-6\beta^{2}\epsilon_{\mathrm{th}}}{1+6\beta^{2}\epsilon_{\mathrm{th}}\left(1-r/r_{\mathrm{c}}\right)}\,.
\end{equation}
The current constraint is $\left|\gamma-1\right|=2.3\times10^{-5}$~\cite{Will:2005va,Shapiro:2004zz,Bertotti:2003rm}.
Setting $r=r_{\mathrm{c}}$, for $F(R)$ gravity with $\beta=-1/\sqrt{6}$, the PPN constraint transforms into the constraint on the thin-shell parameter:
\begin{equation}
    \epsilon_{\mathrm{th},\odot}<2.3\times10^{-5}\,.
\end{equation}
This, together with Eq.~\eqref{eq:power-law_phi_sss}, gives
\begin{equation}
    \label{eq:PPN_constraint}
    \frac{2n-1}{12\beta^{2}\Phi_{\mathrm{c}}}\left(\frac{2\rho_{\Lambda}}{\rho_{\mathrm{out}}}\right)^{2n}\xi <2.3\times10^{-5}\,.
\end{equation}
The potential of the Sun is $\Phi_{\mathrm{c}}\simeq2.12\times10^{-6}$, the mean density of our galaxy is $\rho_{\mathrm{out}}\simeq10^{-24}\,\mathrm{g\cdot cm^{-3}}$, and the dark energy density associated with the cosmological constant is approxmiately $\rho_{\Lambda}\simeq10^{-29}\,\mathrm{g\cdot cm^{-3}}$; therefore, the above equation is an inequality in terms of $\xi$ and $n$.
To break down their degeneracy, we consider their relation at the de Sitter point.
Substituting our model \eqref{eq:power-law_sim} into Eq.~\eqref{eq:vacuum_equation} yields
\begin{equation}
    x_{1}-4+\left(2n+1\right)\left(\frac{2}{x_{1}-2}\right)^{2n-1}2\xi+\left(2n-1\right)\left(\frac{2}{x_{1}-2}\right)^{2n}2\xi=0\,,
\end{equation}
where $x_{1}\equiv R_{1}/\Lambda$.
As $x_{1}=4$ in the $\Lambda$CDM model, we assume that the exact de Sitter solution in our model is the $\Lambda$CDM solution plus a term up to the first order of $\xi$, i.e., $x_{1}\approx4+a\xi$.
Substituting it into the above and keeping terms up to the first order gives $a=-8$.
Hence, we obtain $x_{1}\simeq4-8\xi$.
On the other hand, stability conditions require that the vacuum solution satisfies \eqref{eq:stable_de_Sitter}; thus
\begin{equation}
    \label{eq:power-law_xi_n}
    \xi<\frac{1}{8n^{2}+6n-1}\,.
\end{equation}
Comparing it with Eq.~\eqref{eq:PPN_constraint}, we finally obtain
\begin{equation}
    n>0.95\quad\mathrm{or}\quad n<0.50001\,.
\end{equation}
We choose $n>0.95$ to distinguish the $\Lambda$CDM model explicitly.
Correspondingly, $\xi$ is constrained to $\xi<0.083$.

\subsection{Equivalence principle constraints}

Finally, we constrain our model through tests for violation of the equivalence principle.
We consider the thin-shell effect \eqref{eq:filed_profile} of the Earth, the Moon, and the Sun.
Then, the relevant accelerations are
\begin{align}
    a_{5} & =\left|\kappa\beta_{\mathrm{eff}}\vec{\nabla}\varphi(r)\right|=2\beta_{\mathrm{eff}}\beta_{\mathrm{eff},\odot}\frac{GM_{\odot}}{r^{2}}\,, \\
    a     & =a_{\mathrm{N}}+a_{5}=\frac{GM_{\odot}}{r^{2}}\left(1+2\beta_{\mathrm{eff}}\beta_{\mathrm{eff},\odot}\right)\,,
\end{align}
where the subscripts "5" and "N" denote the fifth force and Newtonian gravity, respectively, and $a_{\mathrm{N}}=\frac{GM_{\odot}}{r^{2}}$.
Therefore, the total accelerations for the Earth ($a_{\oplus}$) and Moon ($a_{\oslash}$) are
\begin{align}
    a_{\oplus}  & =\frac{GM_{\odot}}{r^{2}}\left(1+18\beta^{2}\epsilon_{\mathrm{th},\oplus}\epsilon_{\mathrm{th},\odot}\right)\simeq\frac{GM_{\odot}}{r^{2}}\left(1+18\beta^{2}\epsilon_{\mathrm{th},\oplus}^{2}\frac{\Phi_{\oplus}}{\Phi_{\odot}}\right)\,,                    \\
    a_{\oslash} & =\frac{GM_{\odot}}{r^{2}}\left(1+18\beta^{2}\epsilon_{\mathrm{th},\oslash}\epsilon_{\mathrm{th},\odot}\right)\simeq\frac{GM_{\odot}}{r^{2}}\left[1+18\beta^{2}\epsilon_{\mathrm{th},\oplus}^{2}\frac{\Phi_{\oplus}^{2}}{\Phi_{\oslash}\Phi_{\odot}}\right]\,.
\end{align}
On the other hand, the upper limit on the relative difference in free-fall acceleration toward the Sun, obtained from Lunar Laser Ranging experiments, is given by~\cite{PhysRevLett.83.3585}
\begin{equation}
    2\frac{\left|a_{\oplus}-a_{\oslash}\right|}{\left|a_{\oplus}+a_{\oslash}\right|} <\times10^{-13}\,.
\end{equation}
Substituting the gravitational potentials $\Phi_{\oplus}\simeq7.0\times10^{-10}$, $\Phi_{\oslash}\simeq3.1\times10^{-11}$, and $\Phi_{\odot}\simeq2.1\times10^{-6}$ into the above gives
\begin{equation}
    \epsilon_{\mathrm{th},\oplus} <1.3\times10^{-6}\,.
\end{equation}
According to Eq.~\eqref{eq:thin-shell_parameter}, the field profile outside the Earth is bounded to
\begin{equation}
    \label{eq:EP_phi}
    \left|\kappa\varphi_{\mathrm{out},\oplus}\right| <3.7\times10^{-16}\,.
\end{equation}
The corresponding constraint on the Jordan-frame field amplitude is
\begin{equation}
    \label{eq:EP_fR}
    \left|f_{R,\mathrm{out}}\right| <3.0\times10^{-15}\,.
\end{equation}
Note that the above two constraints are model-independent.
We finally apply the constraints to our model \eqref{eq:power-law_xi_n} and obtain the lower limit of the model parameter
\begin{equation}
    n >1.4\,.
\end{equation}
It is more stringent than that inferred from the post-Newtonian constraint.

\section{Unified description}
\label{sec:unified_description}

In this section, we discuss the unified description of the primordial and late-time acceleration in $F(R)$ gravity, $F(R)=F_{\mathrm{de}}(R)+\frac{R^{2}}{6M^{2}}$.
The first derivative is
\begin{equation}
    F_{R}(R)=F_{R}^{\mathrm{(de)}}(R)+\frac{R}{3M}\,.
\end{equation}
After inflation, the Ricci scalar oscillates around its vacuum state, resulting in explosive particle production.
For model stability, the condition $F_{R}>0$ must be satisfied.
Given that the last term on the right-hand side is negative for $R<0$, $F_{R}^{\mathrm{(de)}}(R)$ must be positive to meet the stability requirement.
Consequently, viable models, such as Hu-Sawicki, Starobinsky, Tsujikawa, and exponential models, do not satisfy this condition.
In contrast, globally stable dark energy models are constructed by requiring $F_{R}^{\mathrm{(de)}}(R)>0$ for all $R$ initially, thus potentially allowing for a unified description.
However, even globally stable models require some modifications to adapt to this unified scenario fully.

In our construction, $F_{R}^{(\mathrm{de})}(R\ll R_{\mathrm{c}})\to0$, which also leads to $F_{R}<0$ for $R<0$.
Fortunately, this issue can be addressed with minor improvements.
To maintain a positive $F_{R}$ within the interval $-M^2\ll R\ll M^2$ of interest, we can adjust the asymptotic behavior of the original dark energy model to offset the negative contribution from the term $\frac{R}{3M^{2}}$.
Simultaneously, to remain consistent with late-time observations, we should retain the original asymptotic behavior $F(R)\to R-2\Lambda$ for $R_{\text{c}}\ll R\ll M^{2}$.
We discuss the Appleby-Battye model and our model separately in the following subsections.

\subsection{Improved Appleby-Battye model}

The improvement of the original Appleby-Battye model has already been studied in Refs.~\cite{Appleby:2009uf, Ribeiro:2023yhh}, where the resulting model is called the $gR^{2}$-AB model and is expressed as
\begin{equation}
    \label{eq:gR2-AB}
    F(R) =\left(1-g\right)R+g\epsilon\ln\left[\frac{\cosh\left(R/\epsilon-b\right)}{\cosh b}\right]+\frac{R^{2}}{6M^{2}}\,,
\end{equation}
where the factor $g$ is introduced to compensate for the negative term $\frac{R}{3M^{2}}$.
It is restricted to $0<g<1/2$; for $g=1/2$ this model becomes the $R^{2}$-AB model, and for $g=0$ it recovers the original $R^{2}$ inflationary model.
This factor is obtained by integrating $F_{RR}$ over $-M^{2}\ll R\ll M^{2}$ and requiring $F_{RR}>0$ \cite{Appleby:2009uf}:
\begin{equation}
    g=\frac{F_{R}(R)-F_{R}(-R)}{2F_{R}(R)}\,,\qquad R_{0}\ll R\ll M^{2}\,.
\end{equation}
We shall derive it from a different point of view.

We require that the $F_{R}(R)$ function behave as a sigmoid curve and the original Appleby-Battye model correspond to the logistic function \eqref{eq:logistic_function}, where the lower limit approaches zero as $R\to -infty$.
However, we aim for $F_{R}$ to decrease more gradually for $-M^{2}<R/R_{\mathrm{c}}\ll-1$, which necessitates a positive lower limit.
To achieve this, we introduce the modified logistic function:
\begin{equation}
    S(x)=B+\frac{L-B}{1+e^{k(x-x_{0})}}\,,
\end{equation}
where $B$, $L$, and $x_{0}$ represent the lower limit, upper limit, and center of the curve, respectively, and $k$ controls the slope of the curve.
Then, the favored $F_R(R)$ function is given by
\begin{equation}
    F_{R}(R)=j+\frac{1-j}{1+be^{-R/R_{\mathrm{c}}}}+\frac{R}{3M^{2}}\,,
\end{equation}
where we introduce a factor $j$, analogous to $g$: $j=0$ corresponds to the $R^{2}-$AB model and $j=1$ to the $R^{2}$ inflation.

Integrating $F_{R}$ yields
\begin{equation}
    F(R)=jR+\left(1-j\right)\ln\left(b+e^{R/R_{\mathrm{c}}}\right)R_{\mathrm{c}}+\frac{R^{2}}{6M^{2}}+C\,.
\end{equation}
where $C$ is an integration constant that can be fixed by imposing $F(0)=0$:
\begin{equation}
    C=\left(j-1\right)\ln\left(1+b\right)R_{\mathrm{c}}\,.
\end{equation}
Thus, the model is
\begin{equation}
    \label{eq:jR2-logistic}
    F(R) =R-\left(1-j\right)R_{\mathrm{c}}\left[\ln\left(1+b\right)-\ln\left(1+be^{-R/R_{\mathrm{c}}}\right)\right]+\frac{R^{2}}{6M^{2}}\,.
\end{equation}
With the following redefinition, one observes that it is equivalent to the $gR^{2}$-AB model \eqref{eq:gR2-AB}:
\begin{align}
    b              & \to e^{2b}\,,                                  \\
    1-j            & \to2g\,,                                       \\
    R_{\mathrm{c}} & \to\frac{\mathrm{\ensuremath{\epsilon}}}{2}\,.
\end{align}

\subsection{Our model}

Finally, we discuss improving our model to realize a unified description.
Following the approach of the previous section, we introduce the $j$-factor:
\begin{equation}
    F_{R}(R)=1-j+j\frac{\sqrt{\left(R/R_{\mathrm{c}}-b\right)^{2}+c}+R/R_{\mathrm{c}}-b}{2\sqrt{\left(R/R_{\mathrm{c}}-b\right)^{2}+c}}+\frac{R}{3M^{2}}\,.
\end{equation}
Here, $j=0$ and $j=1$ correspond to the $R^{2}$ inflationary model and the $R^{2}$-corrected model, respectively.
When $R/R_{\mathrm{c}}\gg1$, we observe
\begin{equation}
    F_{R}(R)\simeq1+\frac{R}{3M^{2}}\,,
\end{equation}
and when $R/R_{\mathrm{c}}\ll-1$,
\begin{align}
    F_{R}(R) & =1-2j+\frac{R}{3M^{2}}\,,        \\
    j        & <\frac{1}{2}+\frac{R}{6M^{2}}\,,
\end{align}
as expected.
To ensure that our model remains positive for $-M^{2}<R<M^{2}$, $j$ should satisfy
\begin{equation}
    0<j<\frac{1}{3}\,.
\end{equation}

Integrating $F_{R}$ yields
\begin{equation}
    F(R)=\left(1-\frac{j}{2}\right)R+\frac{j}{2}R_{\mathrm{c}}\sqrt{\left(R/R_{\mathrm{c}}-b\right)^{2}+c}+\frac{R^{2}}{6M^{2}}+C\,,
\end{equation}
where the integration constant can be determined by setting $F(0)=0$:
\begin{equation}
    C=-\frac{j}{2}R_{\mathrm{c}}\sqrt{b^{2}+c}\,.
\end{equation}
Then, our improved model takes the form
\begin{equation}
    F(R)=\left(1-\frac{j}{2}\right)R+\frac{j}{2}R_{\mathrm{c}}\sqrt{\left(R/R_{\mathrm{c}}-b\right)^{2}+c}-\frac{j}{2}R_{\mathrm{c}}\sqrt{b^{2}+c}+\frac{R^{2}}{6M^{2}}\,.
\end{equation}

In the high-curvature limit, we require that $F(R\gg bR_{\mathrm{c}})\to R-2\Lambda+\frac{R^{2}}{6M^{2}}$.
Therefore, we define
\begin{equation}
    \lambda\equiv\frac{j}{2}\left(b+\sqrt{b^{2}+c}\right)\,,
\end{equation}
so that $\lambda R_{\mathrm{c}}=2\Lambda$.
An alternative form of our model can then be expressed as
\begin{equation}
    F(R)=R-\lambda R_{\mathrm{c}}+\frac{j}{2}R_{\mathrm{c}}\sqrt{\left(R/R_{\mathrm{c}}-b\right)^{2}+\left(2\lambda/j-b\right)^{2}-b^{2}}-\frac{j}{2}R_{\mathrm{c}}\left(R/R_{\mathrm{c}}-b\right)+\frac{R^{2}}{6M^{2}}\,.
\end{equation}

Finally, the generalization to a wider range of powers is straightforward:
\begin{equation}
    F(R)=R-\lambda R_{\mathrm{c}}+\frac{j}{2}R_{\mathrm{c}}\left[\left(R/R_{\mathrm{c}}-b\right)^{2n}+\left(2\lambda/j-b\right)^{2n}-b^{2n}\right]^{\frac{1}{2n}}-\frac{j}{2}R_{\mathrm{c}}\left(R/R_{\mathrm{c}}-b\right)+\frac{R^{2}}{6M^{2}}\,.
\end{equation}
Here, $\lambda$ is defined by
\begin{equation}
    \lambda\equiv\frac{j}{2}\left[b+\left(b^{2n}+c\right)^{1/2n}\right]\,,
\end{equation}
and the parameter $b$ must satisfy
\begin{equation}
    b<\lambda/j\,.
\end{equation}

\section{Conclusion}

In this study, we generalized the approach of Appleby and Battye for developing an alternative framework for dark energy model building in $F(R)$ gravity \cite{Appleby:2007vb}.
As most existing dark energy models are prone to weak curvature singularities \cite{Starobinsky:2007hu, Tsujikawa:2007xu, Frolov:2008uf, Appleby:2008tv}, adding a high-curvature term, such as $\frac{1}{6M^2}R^2$, appears necessary to eliminate the singularity \cite{Kobayashi:2008wc, Appleby:2009uf, Lee:2012dk}.
On the one hand, including an $R^2$ correction preserves the viability of existing dark energy models.
On the other hand, given the success of the $R^2$ inflation \cite{Starobinsky:1980te, Planck:2018jri}, it is appealing to describe primordial and late-time cosmic acceleration within a single $F(R)$ Lagrangian.
In the latter case, a smooth transition between the two acceleration phases is essential for a successful cosmological model.
Specifically, post-inflationary reheating in the original $R^2$ inflation scenario requires an oscillating Ricci scalar around the vacuum state.
However, most dark energy models struggle to satisfy this requirement, as they become unstable when $R < 0$, i.e., $F_{R} < 0$ and/or $F_{RR} < 0$.

To address this issue, we showed that globally stable dark energy models can be constructed by initially imposing $F_{R} > 0$ and $F_{RR} > 0$ for all $R$, following the ideas of Appleby and Battye \cite{Appleby:2007vb}.
Furthermore, we propose that the favored $F_{R}$ function should not only be positive and bounded but also exhibit a sigmoid shape.
Thus, we reformulated and extended the original Appleby-Battye model.
As the Appleby-Battye model includes exponential terms, it shares similarities with the Tsujikawa and exponential models.
For comparison, we introduced a model whose first derivative exhibits a power-law behavior in the high-curvature limit. The model inherits the simplicity of the Hu-Sawicki model and achieves global stability, ensuring compatibility with the $R^2$ inflationary model.

We then explicitly solved the modified Friedmann equation, demonstrating that our model can successfully explain the current cosmic acceleration.
We observed that the EoS for the dark energy component exhibits phantom crossing behavior, a universal feature of $F(R)$ gravity.
Subsequently, we constrained the model parameters by confronting our model with local gravity tests.
Finally, we introduced a $j$ factor to ensure stability throughout cosmic history for unified models.

This study has only focused on model building, background evolution, and local gravity tests.
However, Cosmological tests are essential to validate our model further.
Additionally, a detailed analysis of reheating and a comparison with the original $R^2$ model are required.
These explorations are left for future work.

\bibliographystyle{apsrev4-1}
\bibliography{refs}

\end{document}